# Spin states of Co ions in La$_{1.5}$Ca$_{0.5}$CoO$_4$ from first principles

Ting Jia,[1] Hua Wu,[2,*,†] Guoren Zhang,[1] Xiaoli Zhang,[1] Ying Guo,[1] Zhi Zeng,[1,*,‡] and H. Q. Lin[3]

[1]*Key Laboratory of Materials Physics, Institute of Solid State Physics, Chinese Academy of Sciences, Hefei 230031, People's Republic of China*

[2]*II. Physikalisches Institut, Universität zu Köln, Zülpicher Str. 77, 50937 Köln, Germany*

[3]*Department of Physics and Institute of Theoretical Physics, The Chinese University of Hong Kong, Shatin, Hong Kong, People's Republic of China*

The spin states and electronic structure of layered perovskite La$_{1.5}$Ca$_{0.5}$CoO$_4$ are investigated using full-potential linearized augmented plane-wave method. All the computational results indicate that the Co$^{2+}$ ion is in a high-spin (HS, $t_{2g}^5 e_g^2$) state and the Co$^{3+}$ in a low-spin (LS, $t_{2g}^6$) state. The Co$^{2+}$ $t_{2g}$ orbitals with a small crystal-field splitting are mixed by spin-orbit coupling, which accounts for the observed easy in-plane magnetism. The nonmagnetic LS-Co$^{3+}$ state, which is stabilized by a strong crystal field, provides a natural explanation for the observed low magnetic ordering temperature and a spin-blockade phenomenon of the electron hopping. Furthermore, we find that the intermediate-spin (IS, $t_{2g}^5 e_g^1$) state of Co$^{3+}$ has a large multiplet splitting. But the lowest-lying IS state of Co$^{3+}$ is still higher in energy than the LS ground state by a few hundred millielectron volts and the HS state of Co$^{3+}$ is even less stable, both in sharp contrast to a recent experimental study which suggested the HS+IS mixed Co$^{3+}$ ground state. We note that either the IS-Co$^{3+}$ or HS-Co$^{3+}$ states or their mixture would produce a wrong out-of-plane magnetic anisotropy and a much higher magnetic-ordering temperature than observed. Thus, the present work sheds light on this material concerning its electronic and magnetic structure, and it would stimulate different experiments to settle this intriguing spin-state issue.

## I. INTRODUCTION

Cobaltates have been discovered with properties including superconductivity,[1] giant magnetoresistance,[2–4] and spin blockade.[5] Their interesting properties are produced by the interplay between different degrees of freedom such as spin, charge, and orbital. Such an interplay occurs in many transition-metal oxides.[6] In addition, the spin-state degree of freedom distinguishes cobaltates from other transition-metal oxides. Particularly for the Co$^{3+}$ ions, it can be low spin (LS, $S=0$), intermediate spin (IS, $S=1$), and high spin (HS, $S=2$). These different spin states arise from the competition between crystal-field effects and Hund's exchange interactions. One prototype oxide is the perovskite LaCoO$_3$, in which the crystal-field splitting (CFS) is close to the exchange energy. The ground state of this compound is believed to be a nonmagnetic LS state.[7] Two magnetic transitions at about 100 and 500 K have been attributed to the temperature-induced spin-state transition of Co$^{3+}$ ions, although the spin-state transitions [LS→HS,[8–12] LS→IS,[13–16] or LS→(HS+LS) →IS[17–19]] are still controversial.

Layered perovskites La$_{2−x}$A$_x$CoO$_4$ (A=Sr and Ca) are another group of cobaltates, and they received considerable attention recently due to their interesting structural, electronic, and magnetic properties which are closely linked to the intriguing spin-state issue.[20–30] Such properties of the Sr-doped La$_{1.5}$Sr$_{0.5}$CoO$_4$ system have not been well understood until a very recent group of studies,[24–28] which have settled the spin-state issue and convincingly proven that the charge-ordered high-spin Co$^{2+}$ and low-spin Co$^{3+}$ state is a key. The Ca-doped La$_{1.5}$Ca$_{0.5}$CoO$_4$ has similar physical properties to the above La$_{1.5}$Sr$_{0.5}$CoO$_4$. One might expect the same spin state in both. However, a strong controversy arises, as recent neutron-scattering experiments[29] showed that in La$_{1.5}$Ca$_{0.5}$CoO$_4$ both the Co$^{2+}$ and Co$^{3+}$ ions are in the high-spin state. Furthermore, more recent magnetic-susceptibility measurements and x-ray fluorescence spectroscopy concluded that the Co$^{3+}$ ions in La$_{1.5}$Ca$_{0.5}$CoO$_4$ are rather in the high-spin and intermediate-spin mixed state.[30] Obviously, the spin state in La$_{1.5}$Ca$_{0.5}$CoO$_4$ is an interesting but unsettled issue, and thus we are motivated to carry out detailed first-principles calculations to address this issue. As will be seen below, our study does not support the latest experimental findings on La$_{1.5}$Ca$_{0.5}$CoO$_4$ but reaches basically the same conclusions as very recently made on La$_{1.5}$Sr$_{0.5}$CoO$_4$. Therefore, if our results could not be of a surprise, they indeed confirm theoretically that both systems have the identical spin state and as a result, similar electronic/magnetic structures as both do in reality. It is believed that our study would stimulate different experiments, which could confirm our conclusions and eventually settle the intriguing spin-state issue in La$_{1.5}$Ca$_{0.5}$CoO$_4$.

## II. COMPUTATIONAL DETAILS

We have carried out first-principles calculations to study the charge-ordered La$_{1.5}$Ca$_{0.5}$CoO$_4$, using its structural data measured by single-crystal neutron diffraction,[30] and employing the full-potential augmented plane wave plus local-orbital code WIEN2K.[31] The muffin-tin sphere radii were chosen to be 2.31, 1.86, and 1.65 a.u. for La/Ca, Co, and O atoms, respectively. A virtual atom with an atomic number $Z=56.75$ ($0.75Z_{La}+0.25Z_{Ba}$) is used for the (La$_{1.5}$Ca$_{0.5}$) sites since La and Ca(Ba) ions are in most cases simply electron

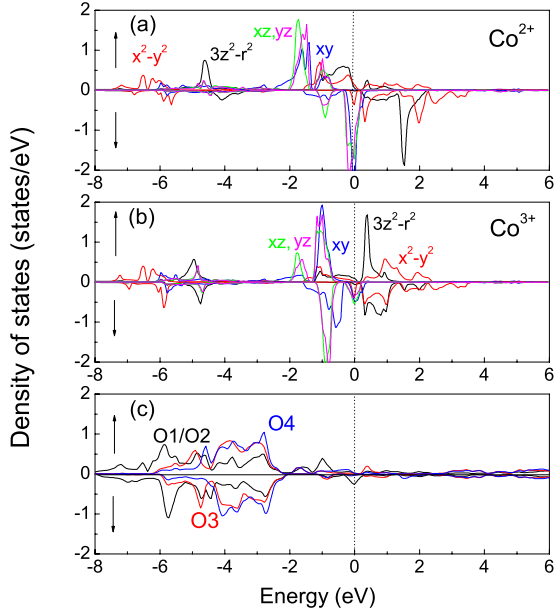

FIG. 1. (Color online) DOS of (a) HS-$Co^{2+}$, (b) LS-$Co^{3+}$, and (c) the planar O1/O2, $Co^{3+}$-apical O3, and $Co^{2+}$-apical O4 in $La_{1.5}Ca_{0.5}CoO_4$ by LSDA.

## III. RESULTS AND DISCUSSIONS

The spin-polarized LSDA calculations, involving the influences of crystal-field effects and magnetic-exchange interactions, show that the electrons of $Co^{3+}$ have been forced into LS state. This indicates that the strong crystal field plays a dominant role in the stabilization of LS state. Especially, such a strong crystal field could originate from a chemical pressure due to the neighboring HS-$Co^{2+}$ with a relatively bigger ionic size. The LSDA results of the ferromagnetic (FM) state are shown in Fig. 1. In the approximate octahedral crystal field, the Co 3*d* orbitals are split into upper $e_g$ and lower $t_{2g}$ states. The more dispersion of the $e_g$ bands both in $Co^{3+}$ and $Co^{2+}$ (see Fig. 1) is due to larger overlap between $e_g$ and O 2*p* orbitals in comparison with that between $t_{2g}$ and O 2*p* orbitals. Under the tetragonal distortion, the triply degenerate $t_{2g}$ orbitals are further split into one $d_{xy}$ and two degenerate $d_{xz}/d_{yz}$ orbitals. At the same time, doubly degenerate $e_g$ orbitals are further split into nondegenerate $d_{x^2-y^2}$ and $d_{3z^2-r^2}$ states. As shown in Fig. 1, the near-degenerate $d_{xz}/d_{yz}$ orbitals lie a little lower than the $d_{xy}$ orbital and the $d_{3z^2-r^2}$ state is lower than the $d_{x^2-y^2}$ orbital both in $Co^{3+}$ and $Co^{2+}$. This corresponds to the z-axis-elongated $CoO_6$ octahedral crystal field.[30] Particularly, the $Co^{2+}$ $t_{2g}$ CFS is estimated to be only about 20 meV, using the center of gravity of the energy for each orbitally resolved density of states. The planar O1/O2 are shared by $Co^{3+}O_6$ and $Co^{2+}O_6$ octahedra in the basal plane while the apical O3 and O4 belong to $Co^{3+}O_6$ and $Co^{2+}O_6$ octahedra, respectively. Figure 1(c) shows spin polarization for O 2*p*. The values of magnetic moment for O1, O2, O3, and O4 atoms are about 0.05 $\mu_B$, 0.05 $\mu_B$, 0.004 $\mu_B$, and 0.05 $\mu_B$, respectively. The appearance of spin polarization and magnetic moment at O1, O2, and O4 sites could be attributed to their strong hybridization with occupied $Co^{2+}$ $d_{x^2-y^2}$ and $d_{3z^2-r^2}$ orbitals while O3 site with little spin polarization and magnetic moment is coupled with the empty $Co^{3+}$ $d_{3z^2-r^2}$ orbital. As shown in Fig. 1(b), the $t_{2g}$ orbitals are almost fully filled and $e_g$ orbitals are nearly empty in $Co^{3+}$, suggesting a LS state of $Co^{3+}(t_{2g}^6 e_g^0)$. Figure 1(a) shows a HS state of $Co^{2+}(t_{2g}^5 e_g^2)$: majority-spin $e_g$ and $t_{2g}$ orbitals are almost filled and minority-spin $t_{2g}$ orbitals are only partially occupied. And the partially filled $t_{2g}$ orbitals cross the Fermi level, suggesting a metallic state in contrast to the experimentally observed insulating behavior. The failure of the LSDA in describing the electronic structure of donors. The cutoff parameter $R_{mt}K_{max}$ was set to 7.0 and 500 *k* points were used over the first Brillouin zone. A supercell having four planar Co ions was used to study the possible combinations of the ordered spin states. Exchange and correlation effects were taken into account in a local-spin-density approximation (LSDA)[32] or generalized gradient approximation (GGA) by Perdew, Burk, and Ernzerhof (PBE).[33] To account for the strong electron correlations, three different schemes were used: the LSDA+*U* method[34] in the so-called "full localized limit,"[35] the GGA+*U* method, and a hybrid functional PBE0.[36] LSDA+*U* or GGA+*U* calculations were performed with $U_{eff}=U-J=4.1$ eV (on-site Coulomb repulsion $U=5$ eV and Hund exchange constant $J=0.9$ eV).[27,37,38] The spin-orbit coupling (SOC) turns out to be important and is included by the second-variational method with scalar relativistic wave functions.[31] The magnetic-exchange interactions, correlation effects, multiplet effect, and the SOC are included gradually to investigate the electronic structure and magnetic properties of $La_{1.5}Ca_{0.5}CoO_4$ as seen below.

TABLE I. The total energies *E* [meV/(2 f.u.)], spin moments *M* ($\mu_B$) of $La_{1.5}Ca_{0.5}CoO_4$ in different spin states calculated by LSDA+*U* in FM coupling. The spin configurations are shown only once. The HS, HS+IS, and HS+LS states of $Co^{3+}$ are unstable and converge to the IS, IS, and IS+LS states, respectively. The LS-$Co^{3+}$/HS-$Co^{2+}$ state is the most stable one and taken as the total-energy reference zero.

| State and configuration | Energy | $Co^{3+}_{spin}$ | $Co^{2+}_{spin}$ |
| --- | --- | --- | --- |
| LS ($t_{2g}^6$) $Co^{3+}$/HS ($t_{2g}^5 e_g^2$)$Co^{2+}$ | 0 | 0.29 | 2.46 |
| IS ($t_{2g}^5 e_g^1$) $Co^{3+}$/HS $Co^{2+}$ | 692 | 1.64 | 2.32 |
| HS ($t_{2g}^4 e_g^2$) $Co^{3+}$/HS $Co^{2+}$ → IS$Co^{3+}$/HS $Co^{2+}$ | 639 | 2.16 | 2.36 |
| HS+IS $Co^{3+}$/HS $Co^{2+}$ → IS$Co^{3+}$/HS $Co^{2+}$ | 520 | 2.14, 2.05 | 2.38 |
| HS+LS $Co^{3+}$/HS $Co^{2+}$ → IS+LS$Co^{3+}$/HS $Co^{2+}$ | 351 | 2.11, 0.31 | 2.41 |

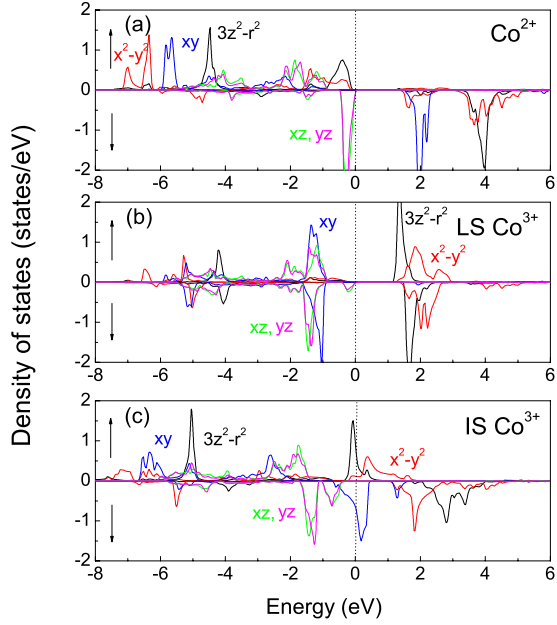

FIG. 2. (Color online) DOS of the (a) HS-Co$^{2+}$ and (b) LS-Co$^{3+}$ ions in LS-Co$^{3+}$/HS-Co$^{2+}$ state, (c) IS-Co$^{3+}$ ions in IS-Co$^{3+}$/HS-Co$^{2+}$ state by LSDA+$U$.

transition-metal oxides is known to be associated with an inadequate description of the strong-correlation effects.

Considering the influence of correlation effects, LSDA+$U$ methods are used to search the ground state of the system, as LSDA+$U$ calculations allow us to access different spin and orbital states by initializing their corresponding density matrix and then doing self-consistently a full electronic relaxation. As shown in Table I, the total-energy results reveal the LS ground state of Co$^{3+}$ ions. The HS-Co$^{3+}$ state is unstable and converges to an IS-Co$^{3+}$ state. Note also that the HS-Co$^{2+}$/LS-Co$^{3+}$ ground state remains to be more stable than HS-Co$^{2+}$/IS-Co$^{3+}$ state by 651(543) meV per Co$^{3+}$ for $U=4$ eV (6 eV). In addition, the GGA+$U$ and PBE0 calculations, confirming the LSDA+$U$ results, show that the LS state is more stable than the IS state by 790 meV and 268 meV per Co$^{3+}$, respectively. Moreover, our results show that only the HS-Co$^{2+}$/LS-Co$^{3+}$ ground state is insulating and has a band gap of 1.3 eV, being consistent with the insulating nature of La$_{1.5}$Ca$_{0.5}$CoO$_4$ from experiments.[29] All other spin states still have a metallic solution when including the correlation effects (but no SOC effect which further lifts the orbital degeneracy, see below the LSDA+$U$+SOC results for the joint effects of the SOC and the Hubbard $U$). As shown in Fig. 2(a), the partially filled $t_{2g}$ orbitals of HS-Co$^{2+}$ ions are split into occupied $d_{xz}/d_{yz}$ orbitals and unoccupied $d_{xy}$ orbital in the down-spin channel. For the LS-Co$^{3+}$ ions [Fig. 2(b)], the correlation effects push the occupied $t_{2g}$ bands down and the unoccupied $e_g$ bands up. The gap of 1.3 eV is thus opened in the HS-Co$^{2+}$/LS-Co$^{3+}$ ground state. However, the IS-Co$^{3+}$ solution is metallic, see Fig. 2(c). The density of states (DOS) of IS-Co$^{3+}$ state has a broader band structure than that of LS-Co$^{3+}$ state. The formally occupied $d_{3r^2-z^2\uparrow}$ and unoccupied $d_{xy\downarrow}$ bands cross the Fermi level.

We go further to include the multiplet effect and SOC, both of which have turned out to be important when studying the ground spin state and understanding the electronic properties,[27,38] LSDA+$U$+SOC calculations are initialized by assuming the different spin and orbital states. As shown in Table II, the LS state of Co$^{3+}$ ions remains to be the ground state. All the calculated spin-state solutions are now insulating with a different size of band gap. In the following, we will analyze all possible spin states one by one to figure out the ground spin state of La$_{1.5}$Ca$_{0.5}$CoO$_4$.

Figure 3 illustrates the DOS of the HS-Co$^{2+}$/LS-Co$^{3+}$ ground state from LSDA+$U$+SOC calculations. It is clear that SOC has no significant influence on the HS-Co$^{2+}$ and LS-Co$^{3+}$ configurations, compared with the LSDA+$U$ results (Fig. 2). However, it is important to note that as the small ionic CFS of the Co$^{2+}$ $t_{2g}$ orbitals is only about 20 meV, the SOC is operative and somehow mixes the higher $d_{xy}$ with lower $d_{xz}/d_{yz}$ orbitals to produce an easy in-plane magnetism. As shown in Table II, The LSDA+$U$+SOC calculations show an easy in-plane magnetism with a relative large orbital moment of 0.33 $\mu_B$/$Co^{2+}$: the state with the magnetic moments in the *ab* plane is more stable than the state with an out-of-plane magnetic direction by 15 meV per $Co^{2+}$. For the LS-$Co^{3+}$, the $t_{2g}$ orbitals are fully occupied and such a closed $t_{2g}^6$ shell produces an almost quenched orbital moment of 0.03 $\mu_B$. Since the HS-$Co^{2+}$ ion has a larger spin moment than LS-$Co^{3+}$ ion, the former would dominate the magnetic anisotropy. Consequently, the magnetic easy axis is within the *ab* plane. This easy in-plane magnetism, resulting from the HS-$Co^{2+}$/LS-$Co^{3+}$ ground-state solution, is in accord with experiments.[30] Moreover, this ground state with a band gap of 1.2 eV is consistent with the insulating nature of $La_{1.5}Ca_{0.5}CoO_4$.[29] Note that an effective moment of 3.96 $\mu_B$, obtained recently from a fitting of the magnetic susceptibility data above 100 K to a Curie-Weiss law, refers to the HS state of the $Co^{2+}$ ions and the HS+IS mixed state of the $Co^{3+}$.[30] However, such a fitting could be questionable as the SOC split multiplets of the HS $Co^{2+}$ ions may get involved upon a thermal population. But a spin-state transition of the $Co^{3+}$ ions from the LS ground state to either IS or HS state is quite unlikely, as the corresponding excitation energy is as high as several hundred millielectron volts, see Tables I and II. In this sense, further studies should be carried out to settle this intriguing spin-state issue.

TABLE II. The total energies $E$ [meV/(2 f.u.)], the spin and orbital moments $M$ ($\mu_B$) of La$_{1.5}$Ca$_{0.5}$CoO$_4$ in different spin states calculated by LSDA+$U$+SOC. The HS-Co$^{2+}$ configuration is shown only once. Only the ground state has an easy in-plane magnetism (the spin and orbital moments marked by a subscript "$ab$"). All other states have a wrong out-of-plane magnetism. The LS-Co$^{3+}$/HS-Co$^{2+}$ state is the most stable one and taken as the total-energy reference zero.

| State and configuration | Energy | Co$^{3+}_{spin}$ | Co$^{3+}_{orb}$ | Co$^{2+}_{spin}$ | Co$^{2+}_{orb}$ |
|---|---|---|---|---|---|
| LS ($t^6_{2g}$) Co$^{3+}$/HS ($t^3_{2g\uparrow}e^2_{g\uparrow}xz^1_\downarrow yz^1_\downarrow$) Co$^{2+}$ | 0 | 0.28$_{ab}$ | 0.03$_{ab}$ | 2.46$_{ab}$ | 0.33$_{ab}$ |
| LS ($t^6_{2g}$)Co$^{3+}$/HS Co$^{2+}$ | 15 | 0.28 | 0.03 | 2.46 | 0.01 |
| IS1 Co$^{3+}$ [$t^3_{2g\downarrow}(3z^2-r^2)^1_\downarrow xz^1_\uparrow yz^1_\uparrow$]/HS Co$^{2+}$ | 888 | −1.50 | 0.01 | 2.45 | 0.02 |
| IS2 Co$^{3+}$ [$t^3_{2g\downarrow}(3z^2-r^2)^1_\downarrow xy^1_\uparrow (xz-iyz)^1_\uparrow$]/HS Co$^{2+}$ | 520 | −1.49 | −0.78 | 2.47 | 0.01 |
| IS3 Co$^{3+}$ [$t^3_{2g\uparrow}(3z^2-r^2)^1_\uparrow xy^1_\downarrow (xz+iyz)^1_\downarrow$]/HS Co$^{2+}$ | 332 | 2.00 | 0.84 | 2.42 | 0.02 |
| HS1 Co$^{3+}$ [$t^3_{2g\downarrow}e^2_{g\downarrow}(xz-iyz)^1_\uparrow$]/HS Co$^{2+}$ | 461 | −2.93 | −0.81 | 2.42 | 0.01 |
| HS2 Co$^{3+}$ [$t^3_{2g\uparrow}e^2_{g\uparrow}(xz+iyz)^1_\downarrow$]/HS Co$^{2+}$ →IS3 Co$^{3+}$/HS Co$^{2+}$ | 330 | 2.00 | 0.84 | 2.42 | 0.02 |
| HS1+IS3 Co$^{3+}$/HS Co$^{2+}$ | 397 | −2.93, 1.98 | −0.81, 0.84 | 2.42 | 0.02 |
| HS1+LS Co$^{3+}$/HS Co$^{2+}$ | 239 | −2.93, 0.29 | −0.81, 0.03 | 2.43 | 0.02 |

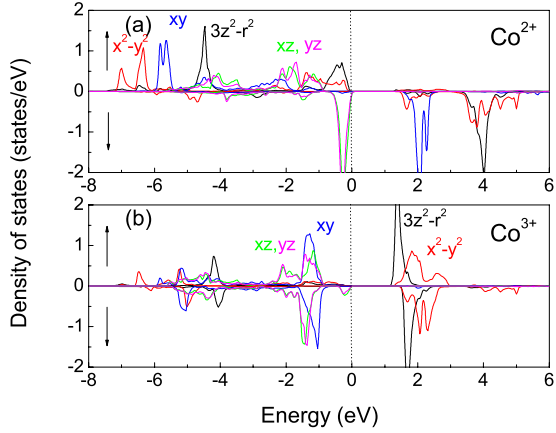

FIG. 3. (Color online) DOS of the (a) HS-$Co^{2+}$/(b) LS-$Co^{3+}$ ground state of $La_{1.5}Ca_{0.5}CoO_4$ by LSDA+$U$+SOC.

We now turn to the IS state of $Co^{3+}$ ions (Fig. 4). In a *z*-axis-elongated $Co^{3+}O_6$ octahedron of $La_{1.5}Ca_{0.5}CoO_4$, the IS-$Co^{3+}$ state configuration $t_{2g\uparrow}^3(3z^2-r^2)_\uparrow^1 xz_\downarrow^1 yz_\downarrow^1$ would be expected, according to a single-electron picture as used quite often in the literature. Note, however, that the occupied $d_{3z^2-r^2}$ electron density has a bigger overlap and thus a stronger Coulomb repulsion with $d_{xz}/d_{yz}$ than with $d_{xy}$. Hence the $d_{xy}$ may be lowered in energy than $d_{xz}/d_{yz}$. Considering such multiplet effect as well as SOC, the IS2 state of $t_{2g\downarrow}^3(3z^2-r^2)_\downarrow^1 xy_\uparrow^1(xz-iyz)_\uparrow^1$ [Fig. 4(b)] configuration [antiferromagnetically (AF) coupled with HS-$Co^{2+}$] is indeed lower in energy than IS1 state of $t_{2g\downarrow}^3(3z^2-r^2)_\downarrow^1 xz_\uparrow^1 yz_\uparrow^1$ [Fig. 4(a)] configuration by 368 meV/$Co^{3+}$. Therefore, the multiplet splitting of $La_{1.5}Ca_{0.5}CoO_4$ is as strong as that of $La_{1.5}Sr_{0.5}CoO_4$ (430 meV).[27] Moreover, the empty $d_{x^2-y^2}$ orbital of $Co^{3+}$ ion is likely to be FM coupled with the half-filled $d_{x^2-y^2}$ orbital of $Co^{2+}$ ion, according to the Goodenough-Kanamori-Anderson (GKA) rules.[39] Indeed, our results (Table II) reveal that the IS3 state of $t_{2g\uparrow}^3(3z^2-r^2)_\uparrow^1 xy_\downarrow^1(xz+iyz)_\downarrow^1$ [Fig. 4(c)] configuration (FM coupled with HS-$Co^{2+}$) is more stable than the AF coupled one by 188 meV/$Co^{3+}$. Such a strong ferromagnetism is in disagreement with the observed low magnetic ordering temperature $T_N$=50 K.[29] Furthermore, the lowest IS3 state out of its multiplet is still higher in energy than the LS ground state by 332 meV/$Co^{3+}$. The IS3 state would have a large out-of-plane orbital moment of 0.84 $\mu_B$ [mainly from the occupied complex orbital $d_{xz+iyz}$ ($d_1$), Fig. 4(c)], which is again in disagreement with the observed easy in-plane magnetism. Finally, the HS-$Co^{2+}$/IS-$Co^{3+}$ solutions are all insulating with band gaps of 0.03 eV for IS1, 0.56 eV for IS2, and 0.30 eV for IS3 states (Fig. 4), all of which seem too small to coincide with the experiments.

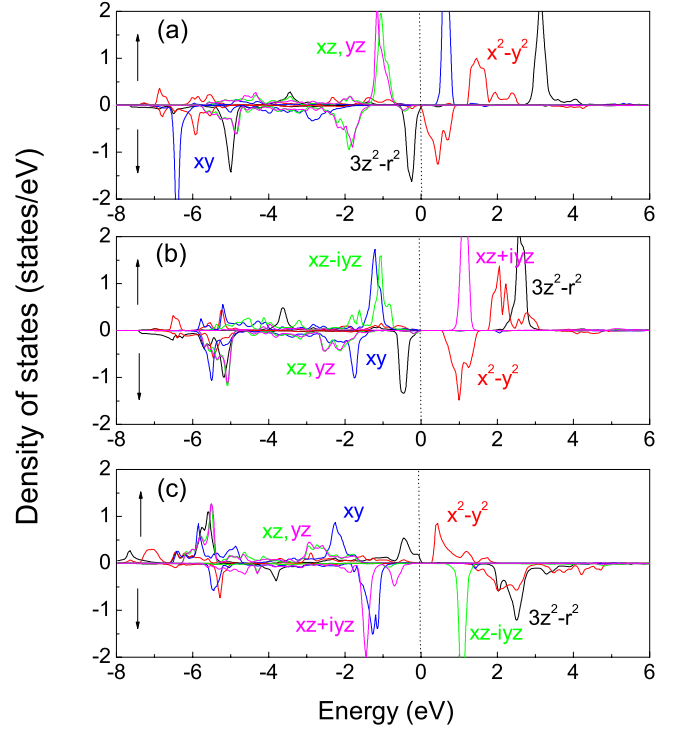

FIG. 4. (Color online) DOS of the IS-$Co^{3+}$ ions in different spin configurations: (a) $t_{2g\downarrow}^3(3z^2-r^2)_\uparrow^1 xz_\uparrow^1 yz_\uparrow^1$, (b) $t_{2g\downarrow}^3(3z^2-r^2)_\downarrow^1 xy_\uparrow^1(xz-iyz)_\uparrow^1$, and (c) $t_{2g\uparrow}^3(3z^2-r^2)_\uparrow^1 xy_\downarrow^1(xz+iyz)_\downarrow^1$ by LSDA+$U$+SOC. The DOS of HS-$Co^{2+}$ ions are not shown here but can refer to Fig. 3.

For the case of HS-$Co^{2+}$/HS-$Co^{3+}$ state which was used to explain the large effective moment of 3.96 $\mu_B$,[30] the ferromagnetically coupled HS-$Co^{2+}$/HS-$Co^{3+}$ state is unstable and converges to a HS-$Co^{2+}$/IS-$Co^{3+}$ state in our calculations (Table II). According to the GKA rules,[39] both HS-$Co^{2+}$ and HS-$Co^{3+}$ ions having a half-filled $d_{x^2-y^2}$ orbital prefer a strong AF coupling, which would yield a high magnetic ordering temperature as in the parent compound $La_2CoO_4$ having the HS $Co^{2+}$ ions and a quite high $T_N$=275 K.[20] Reversely, the low $T_N$=50 K of $La_{1.5}Ca_{0.5}CoO_4$ just infers that the $Co^{3+}$ ions are unlikely in the HS state. This inference is supported by our results that the possible AF-coupled HS-$Co^{2+}$/HS-$Co^{3+}$ state is much higher in energy than the HS-$Co^{2+}$/LS-$Co^{3+}$ ground state by 461 meV/$Co^{3+}$ (see Table II). Moreover, the HS-$Co^{2+}$/HS-$Co^{3+}$ state is quite un-likely because of the large out-of-plane orbital moment of the HS-Co$^{3+}$ ($-0.81$ $\mu_B$): the HS-Co$^{3+}$ ions would dominate the magnetic anisotropy and lead to an easy out-of-plane magnetism, which is in disagreement with the observed easy in-plane magnetism.

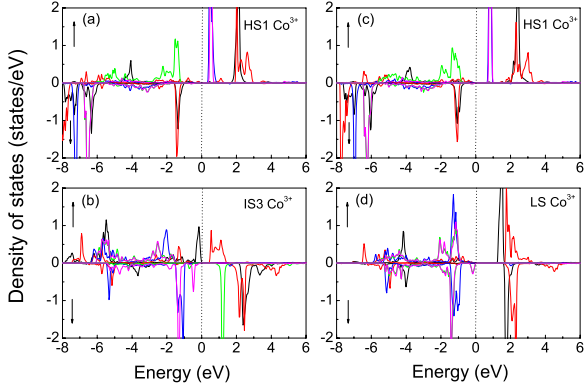

FIG. 5. (Color online) DOS of the (a) HS1-Co$^{3+}$ and (b) IS3-Co$^{3+}$ ions in HS1+IS3-Co$^{3+}$/HS-Co$^{2+}$ state, (c) HS1-Co$^{3+}$ and (d) LS-Co$^{3+}$ ions in HS1+LS-Co$^{3+}$/HS-Co$^{2+}$ state by LSDA+$U$+SOC. The different colors stand for five $d$ orbitals as shown in Figs. 3 and 4. The DOS of HS-Co$^{2+}$ ions are not shown here but can refer to Fig. 3.

About the mixed HS-Co$^{2+}$/HS+IS-Co$^{3+}$ and HS-Co$^{2+}$/HS+LS-Co$^{3+}$ states, within the most possible configurations, HS1-Co$^{3+}$ and IS3-Co$^{3+}$ ions are AF and FM coupled with HS-Co$^{2+}$ ions, respectively. The HS1-Co$^{3+}$ states with one spin-up occupied orbital [$d_{xz\text{-}iyz}(d_{-1})$] [Figs. 5(a) and 5(c)] have large out-of-plane orbital moment of $-0.81$ $\mu_B$ both in HS+IS and HS+LS states. And the DOS of IS3- [Fig. 5(b)] and LS- [Fig. 5(d)] Co$^{3+}$ ions are not significantly changed, comparing with that of pure IS3-Co$^{3+}$ and LS-Co$^{3+}$ ions. The insulating gaps of 0.4 eV for HS1+IS3 and 0.7 eV for HS1+LS are smaller than LS state. Moreover, the HS+LS state lies lower in energy than HS+IS state but still higher in energy than pure LS state (Table II). There is also no any evidence indicating unequal distortions on the Co$^{3+}$ sites due to the different size of the HS-Co$^{3+}$, IS-Co$^{3+}$, and LS-Co$^{3+}$ ions. Thus, the present results reveal that the mixed HS+IS state is not the ground state as previously suggested.[30]

So far, all the possible spin states of Co$^{3+}$ have been discussed. Either the HS or IS state is not likely to exist in the system: neither the magnetic anisotropy nor the magnetic-ordering temperature of La$_{1.5}$Ca$_{0.5}$CoO$_4$ can be explained by these states. Instead, our results have shown that only the LS state could provide a natural explanation for those experimental results.

Besides the HS-Co$^{2+}$/LS-Co$^{3+}$ ground state, La$_{1.5}$Ca$_{0.5}$CoO$_4$ has similar properties with its sister compound La$_{1.5}$Sr$_{0.5}$CoO$_4$: (1) Upon the Ca or Sr doping, the presence of LS-Co$^{3+}$ ions results in a rapid lowering of the $T_N$, which drops from 275 K for La$_2$CoO$_4$ to only 50 K for La$_{1.5}$Ca$_{0.5}$CoO$_4$ and 30 K for La$_{1.5}$Sr$_{0.5}$CoO$_4$. The relatively higher $T_N$ of La$_{1.5}$Ca$_{0.5}$CoO$_4$ can be explained by its shorter Co-O-Co bond length (due to the smaller ionic size of Ca) than that of La$_{1.5}$Sr$_{0.5}$CoO$_4$. (2) The easy in-plane magnetism of both systems originates from the magnetic anisotropy of HS-Co$^{2+}$ ions. (3) Their insulating behavior can be explained by the LS-Co$^{3+}$ ions in between HS-Co$^{2+}$ ions. Considering a pair of neighboring HS-Co$^{2+}$ ($S=3/2$) and LS-Co$^{3+}$ ($S=0$) ions, the hopping of one electron will produce a different IS-Co$^{3+}$ ($S=1$) and LS Co$^{2+}$ ($S=1/2$) final state. As a result, the hopping will be significantly suppressed due to a large energy cost associated with the change in the spin states,[24] being referred to as a spin blockade.[5] Overall, the HS-Co$^{2+}$/LS-Co$^{3+}$ ground state would be a right picture for La$_{1.5}$Ca$_{0.5}$CoO$_4$, which can successfully explain several experimental results as seen above.

## IV. CONCLUSIONS

In summary, we have investigated the spin states and electronic structure of La$_{1.5}$Ca$_{0.5}$CoO$_4$ by first-principles calculations. All the obtained results conclude the HS-Co$^{2+}$/LS-Co$^{3+}$ ground state. The spin-polarized LSDA calculations, involving the influences of crystal-field effects and magnetic exchange interactions, show that the electrons of Co$^{3+}$ ions have been forced into the LS state. This indicates that the strong crystal field plays a dominant role in stabilization of the LS state. The LSDA+$U$ calculations, considering the influence of correlation effects, confirm the HS-Co$^{2+}$/LS-Co$^{3+}$ ground state and provide a good description of its insulating behavior. The LSDA+$U$+SOC calculations, including the SOC and multiplet effect, confirm again the HS-Co$^{2+}$/LS-Co$^{3+}$ ground state and provide a detailed description of the spin configurations. Furthermore, the SOC plays an important role in determining the easy magnetism and the comprehensive electronic structure. And the multiplet effect is so strong that one should take it into account when studying the spin state issue of cobaltates.

Our results also indicate that the HS-Co$^{2+}$/LS-Co$^{3+}$ ground state provides a good explanation for several experiments. The Co$^{2+}$ $t_{2g}$ orbitals with small CFS are mixed by SOC, which accounts for the observed easy in-plane magnetism. However, the HS, IS, or mixed spin states of Co$^{3+}$ would result in a wrong out-of-plane magnetism. In addition, the HS- or IS-Co$^{3+}$ ions would be AF or FM coupled strongly with the HS-Co$^{2+}$ ions, both in disagreement with the low magnetic ordering temperature ($T_N=50$ K) of La$_{1.5}$Ca$_{0.5}$CoO$_4$. Finally, the spin-blockade mechanism, being active in the HS-Co$^{2+}$/LS-Co$^{3+}$ ground state, provides an explanation of the insulating nature of La$_{1.5}$Ca$_{0.5}$CoO$_4$.

## ACKNOWLEDGMENTS

This work was supported by the National Science Foundation of China under Grant No. 10504036, the special Funds for Major State Basic Research Project of China(973) under Grant No. 2007CB925004, 863 Project, Knowledge Innovation Program of Chinese Academy of Sciences under Grant No. KJCX2-YW-W07, Director Grants of CASHIPS and CUHK under Grant No. 3110023. The calculations were performed in Center for Computational Science of CASHIPS. H.W. was supported by the Deutsche Forschungsgemeinschaft through SFB 608.

*Corresponding author.
†wu@ph2.uni-koeln.de
‡zzeng@theory.issp.ac.cn

1 K. Takada, H. Sakurai, E. Takayama-Muromachi, F. Izumi, R. A. Dilanian, and T. Sasaki, Nature (London) **422**, 53 (2003).
2 G. Briceño, H. Chang, X. Sun, P. G. Schultz, and X.-D. Xiang, Science **270**, 273 (1995).
3 I. O. Troyanchuk, N. V. Kasper, D. D. Khalyavin, H. Szymczak, R. Szymczak, and M. Baran, Phys. Rev. Lett. **80**, 3380 (1998).
4 C. Martin, A. Maignan, D. Pelloquin, N. Nguyen, and B. Raveau, Appl. Phys. Lett. **71**, 1421 (1997).
5 A. Maignan, V. Caignaert, B. Raveau, D. Khomskii, and G. Sawatzky, Phys. Rev. Lett. **93**, 026401 (2004).
6 Y. Tokura and N. Nagaosa, Science **288**, 462 (2000).
7 P. M. Raccah and J. B. Goodenough, Phys. Rev. **155**, 932 (1967).
8 M. Zhuang, W. Zhang, and N. Ming, Phys. Rev. B **57**, 10705 (1998).
9 S. Noguchi, S. Kawamata, K. Okuda, H. Nojiri, and M. Motokawa, Phys. Rev. B **66**, 094404 (2002).
10 Z. Ropka and R. J. Radwanski, Phys. Rev. B **67**, 172401 (2003).
11 M. W. Haverkort, Z. Hu, J. C. Cezar, T. Burnus, H. Hartmann, M. Reuther, C. Zobel, T. Lorenz, A. Tanaka, N. B. Brookes, H. H. Hsieh, H.-J. Lin, C. T. Chen, and L. H. Tjeng, Phys. Rev. Lett. **97**, 176405 (2006).
12 A. Podlesnyak, S. Streule, J. Mesot, M. Medarde, E. Pomjakushina, K. Conder, A. Tanaka, M. W. Haverkort, and D. I. Khomskii, Phys. Rev. Lett. **97**, 247208 (2006).
13 M. A. Korotin, S. Yu. Ezhov, I. V. Solovyev, V. I. Anisimov, D. I. Khomskii, and G. A. Sawatzky, Phys. Rev. B **54**, 5309 (1996).
14 J.-Q. Yan, J.-S. Zhou, and J. B. Goodenough, Phys. Rev. B **69**, 134409 (2004).
15 R. F. Klie, J. C. Zheng, Y. Zhu, M. Varela, J. Wu, and C. Leighton, Phys. Rev. Lett. **99**, 047203 (2007).
16 J. M. Rondinelli and N. A. Spaldin, Phys. Rev. B **79**, 054409 (2009).
17 T. Kyômen, Y. Asaka, and M. Itoh, Phys. Rev. B **71**, 024418 (2005).
18 M. Tachibana, T. Yoshida, H. Kawaji, T. Atake, and E. Takayama-Muromachi, Phys. Rev. B **77**, 094402 (2008).
19 K. Knížek, Z. Jirák, J. Hejtmánek, P. Novák, and W. Ku, Phys. Rev. B **79**, 014430 (2009).
20 K. Yamada, M. Matsuda, Y. Endoh, B. Keimer, R. J. Birgeneau, S. Onodera, J. Mizusaki, T. Matsuura, and G. Shirane, Phys. Rev. B **39**, 2336 (1989).
21 Y. Moritomo, K. Higashi, K. Matsuda, and A. Nakamura, Phys. Rev. B **55**, R14725 (1997).
22 I. A. Zaliznyak, J. P. Hill, J. M. Tranquada, R. Erwin, and Y. Moritomo, Phys. Rev. Lett. **85**, 4353 (2000).
23 I. A. Zaliznyak, J. M. Tranquada, R. Erwin, and Y. Moritomo, Phys. Rev. B **64**, 195117 (2001).
24 C. F. Chang, Z. Hu, H. Wu, T. Burnus, N. Hollmann, M. Benomar, T. Lorenz, A. Tanaka, H.-J. Lin, H. H. Hsieh, C. T. Chen, and L. H. Tjeng, Phys. Rev. Lett. **102**, 116401 (2009).
25 M. Cwik, M. Benomar, T. Finger, Y. Sidis, D. Senff, M. Reuther, T. Lorenz, and M. Braden, Phys. Rev. Lett. **102**, 057201 (2009).
26 N. Hollmann, M. W. Haverkort, M. Cwik, M. Benomar, M. Reuther, A. Tanaka, and T. Lorenz, New J. Phys. **10**, 023018 (2008).
27 H. Wu and T. Burnus, Phys. Rev. B **80**, 081105(R) (2009).
28 L. M. Helme, A. T. Boothroyd, R. Coldea, D. Prabhakaran, C. D. Frost, D. A. Keen, L. P. Regnault, P. G. Freeman, M. Enderle, and J. Kulda, Phys. Rev. B **80**, 134414 (2009).
29 K. Horigane, H. Hiraka, T. Uchida, K. Yamada, and J. Akimitsu, J. Phys. Soc. Jpn. **76**, 114715 (2007).
30 K. Horigane, H. Nakao, Y. Kousaka, T. Murata, Y. Noda, Y. Murakami, and J. Akimitsu, J. Phys. Soc. Jpn. **77**, 044601 (2008).
31 P. Blaha, K. Schwarz, G. Madsen, D. Kvasnicka, and J. Luitz, WIEN2K package, http://www.wien2k.at
32 V. I. Anisimov, J. Zaanen, and O. K. Andersen, Phys. Rev. B **44**, 943 (1991).
33 J. P. Perdew, K. Burke, and M. Ernzerhof, Phys. Rev. Lett. **77**, 3865 (1996).
34 V. I. Anisimov, I. V. Solovyev, M. A. Korotin, M. T. Czyzyk, and G. A. Sawatzky, Phys. Rev. B **48**, 16929 (1993).
35 A. G. Petukhov, I. I. Mazin, L. Chioncel, and A. I. Lichtenstein, Phys. Rev. B **67**, 153106 (2003).
36 C. Adamo and V. Baronea, J. Chem. Phys. **110**, 6158 (1999).
37 H. Wu, M. W. Haverkort, Z. Hu, D. I. Khomskii, and L. H. Tjeng, Phys. Rev. Lett. **95**, 186401 (2005).
38 H. Wu, Phys. Rev. B **81**, 115127 (2010).
39 J. B. Goodenough, *Magnetism and the Chemical Bond* (Interscience, New York, 1963); J. Kanamori, J. Phys. Chem. Solids **10**, 87 (1959); P. W. Anderson, Phys. Rev. **79**, 350 (1950).